\documentclass[12pt]{article}
\usepackage{amssymb}
\date{}
\newcommand{\bee}{\begin{eqnarray}}
\newcommand{\ede}{\end{eqnarray}}
\begin{document}
\baselineskip=14 pt
\begin{center}
{\Large\bf Tortoise coordinate and Hawking effect in the Kinnersley
spacetime }
\end{center}
\begin{center}
Jian Yang${}^{a,}$\footnote{e-mail : yjccnuphy@yahoo.com.cn} Zheng
Zhao ${}^{b,}$\footnote{e-mail : zhaoz43@hotmail.com}  Wenbiao Liu
${}^{b,}$\footnote{e-mail : wbliu@bnu.edu.cn}
\end{center}
\begin{center}
a, School of Science,\\
Beijing University of Posts and Telecommunications,\\
 Beijing, 100876, China, \\
 b, Department of Physics, Institute of theoretical physics,\\
Beijing Normal University,\\
 Beijing, 100875, China, \\
\end{center}
\begin{abstract}
Hawking effect from the Kinnersley spacetime is investigated using
the improved Damour-Ruffini method with a new coordinate
transformation. Hawking temperature of the horizons can be obtained
point by point. It is found that Hawking temperatures of different
points on the horizons are different. Especially, Hawking
temperature of Rindler horizon is investigated. The touch between a
Kinnersley black hole and its Rindler horizon is considered, and it
shows that the phenomenon is related to the third law of
thermodynamics.
\end{abstract}



\section{Introduction}
It is well known that Hawking effect in a black hole is one of the
most striking phenomena\cite{1,2}. A black hole was found to have
thermal property right after the four laws of black hole
thermodynamics had been built successfully and Hawking radiation had
been discovered. In 1976, Damour and Ruffini proposed a new method
with which one can also calculate Hawking radiation\cite{3}. Using
this method, Liu et al proved that a Kerr-Newman black hole radiates
Dirac particles\cite{4,5}. In 1990's, Z Zhao, X. X. Dai and Z. Q.
Luo improved Damour-Ruffini method to study Hawking effect from some
dynamical black holes. They only investigated several kinds of
dynamical spherically symmetric black holes via the improved
method\cite{6,7}.

Some rapid progress has been made on Hawking effect of dynamical
black holes in a recent several years~\cite{8,9}. J. L. Huang, et al
investigated Hawking effect of a Vaidya black hole using null
geodesic method~\cite{10}. X. M. Liu, et al studied it via
gravitational anomaly method~\cite{11}. Using a new tortoise
coordinate, Hawking effect of some dynamical spherically symmetric
black holes has been investigated in Ref\cite{12}. We got more
accurate surface gravity and Hawking temperature.

It is well known that a stationary black hole has the same
temperature on the horizon. A dynamical spherically symmetric black
hole has also the same temperature on its horizon at the same time
while its temperature varies with time. However, the temperature at
different points on the horizon of an axisymmetric dynamical black
hole may probably different. a variable depending on points. Using a
new tortoise coordinate, Hawking effect of the Kinnersley spacetime
is studied.

The organization of this paper is as follows. In Sec. 2, we will
give a brief overview on the Kinnersley spacetime. In Sec. 3, we
discuss Hawking effect in the Kinnersley black hole under the new
tortoise coordinate transformation. In Sec. 4, the touch between a
Kinnersley black hole and its Rindler horizon is studied. The
conclusion and discussion are given in the last section.

\section{The Kinnersley spacetime
}

The line element of the Kinnersley spacetime is expressed in
Eddington-Finkelstein advanced time coordinate as \cite{13}:
\begin{eqnarray}
& &
ds^{2}=[1-2ar\cos\theta-r^{2}(f^{2}+h^{2}\sin^{2}\theta)-\frac{2m}{r}]dv^{2}-2dvdr \nonumber\\
& & -2r^{2}fdvd\theta-2r^{2}hsin^{2}\theta
dvd\varphi-r^{2}d\theta^{2}-r^{2}sin^{2}\theta d\varphi^{2},
\end{eqnarray}
where
\begin{eqnarray*}
& & f=-a(v)sin\theta+b(v)\sin\varphi+c(v)\cos\varphi,\nonumber\\
& & h=b(v)\cot\theta\cos\varphi-c(v)\cot\theta\sin\varphi,m=m(v).
\end{eqnarray*}
For the case of accelerated rectilinear motion, we have
\begin{displaymath}
b(v)=c(v)=0,
\end{displaymath}
so the line element can be rewritten as
\begin{eqnarray}
& &
ds^{2}=[1-2ar\cos\theta-r^{2}a^{2}\sin^{2}\theta-\frac{2m}{r}]dv^{2}-2dvdr \nonumber\\
& & +2r^{2}a\sin\theta dvd\theta-r^{2}d\theta^{2}-r^{2}sin^{2}\theta
d\varphi^{2}.
\end{eqnarray}

The null hypersurface condition
\begin{equation}
g^{\mu\nu}\frac{\partial f}{\partial x^{\mu}}\frac{\partial
f}{\partial x^{\nu}}=0
\end{equation}
can be rewritten as
\begin{equation}
2\dot{r}-(1-2ar\cos\theta-\frac{2m}{r})+2ar'\sin\theta-\frac{r'^{2}}{r^{2}}=0,\label{c1}
\end{equation}
where $\dot{r}=\frac{\partial r}{\partial v},r'=\frac{\partial
r}{\partial\theta}.$ The Eq.(\ref{c1}) determines the local event
horizon of the Kinnersley spacetime.
\section{Hawking effect from the black hole horizon}
The Klein-Gordon equation in the Kinnersley spacetime is
\begin{eqnarray}
& & 2\frac{\partial^{2}\Phi}{\partial v\partial r}+\frac{2}{r}\frac{\partial\Phi}{\partial v}+(1-2ar\cos\theta-\frac{2m}{r})
\frac{\partial^{2}\Phi}{\partial r^{2}}+2a\sin\theta\frac{\partial^{2}\Phi}{\partial \theta\partial r}\nonumber\\
& &  +\frac{2r-6ar^{2}\cos\theta-2m}{r^{2}}\frac{\partial\Phi}{\partial r}+
\frac{2a\sin\theta}{r}\frac{\partial\Phi}{\partial \theta}+2a\cos\theta\frac{\partial\Phi}{\partial r}
+\frac{1}{r^{2}}\frac{\partial^{2}\Phi}{\partial \theta^{2}}\nonumber\\
& & +\frac{\cot\theta}{r^{2}}\frac{\partial\Phi}{\partial
\theta}+\frac{1}{\sin^{2}\theta}\frac{\partial^{2}\Phi}{\partial
\varphi^{2}}-\mu^{2}\Phi=0,\label{c3}
\end{eqnarray}
where $\mu$ is the mass of a Klein-Gordon particle.

It is well known that when Hawking effect from a Schwarzschild black
hole is investigated using Damour-Ruffini method, the tortoise
coordinate is defined as following\cite{3}
\begin{equation}
r_{\ast}=r+2M\ln[\frac{r-2M}{2M}].
\end{equation}
In the Kinnersley case, we suggest a new tortoise coordinate
transformation as
\begin{eqnarray}
& &
r_{\ast}=r+\frac{1}{2\kappa(v_{0},\theta_{0})}\ln[\frac{r-r_{H}(v,\theta)}{r_{H}(v,\theta)}],
\nonumber\\
& &v_{\ast}=v-v_{0}, \theta_{\ast}=\theta-\theta_{0},
\end{eqnarray}
where both $v_{0}$ and $\theta_{0}$ are constants under tortoise
coordinate transformation. At the same time, $v_{0}$ is the moment
when the particle escapes from the black hole horizon and depicts
evolution of black hole, $\theta_{0}$ is the location where the
particle escapes from the event horizon of black hole and depicts
shape of black hole. According to the new tortoise coordinate
transformation, we have
\begin{eqnarray*}
& &  \frac{\partial}{\partial
r}=[1+\frac{1}{2\kappa(r-r_{H})}]\frac{\partial}{\partial
r_{\ast}},\\
& &  \frac{\partial}{\partial v}=\frac{\partial}{\partial
v_{\ast}}-\frac{r\dot{r}_{H}}{2\kappa
r_{H}(r-r_{H})}\frac{\partial}{\partial
r_{\ast}},\\
& &  \frac{\partial}{\partial \theta}=\frac{\partial}{\partial
\theta_{\ast}}-\frac{rr_{H}'}{2\kappa
r_{H}(r-r_{H})}\frac{\partial}{\partial r_{\ast}},
\end{eqnarray*}
\begin{eqnarray*}
& & \frac{\partial^{2}}{\partial
r^{2}}=[1+\frac{1}{2\kappa(r-r_{H})}]^{2}\frac{\partial^{2}}{\partial
r_{\ast}^{2}}-\frac{1}{2\kappa(r-r_{H})^{2}}\frac{\partial}{\partial
r_{\ast}},\\
& & \frac{\partial^{2}}{\partial r \partial
v}=[1+\frac{1}{2\kappa(r-r_{H})}]\frac{\partial^{2}}{\partial
r_{\ast}
\partial v_{\ast}}+\frac{\dot{r}_{H}}{2\kappa(r-r_{H})^{2}}\frac{\partial}{\partial
r_{\ast}}\\
& & -\frac{r\dot{r}_{H}}{2\kappa
r_{H}(r-r_{H})}[1+\frac{1}{2\kappa(r-r_{H})}]
\frac{\partial^{2}}{\partial r_{\ast}^{2}},
\end{eqnarray*}
\begin{eqnarray*}
& & \frac{\partial^{2}}{\partial r \partial
\theta}=[1+\frac{1}{2\kappa(r-r_{H})}]\frac{\partial^{2}}{\partial
r_{\ast}
\partial \theta_{\ast}}+\frac{r_{H}'}{2\kappa(r-r_{H})^{2}}\frac{\partial}{\partial
r_{\ast}}\nonumber\\
& & -\frac{rr_{H}'}{2\kappa
r_{H}(r-r_{H})}[1+\frac{1}{2\kappa(r-r_{H})}]
\frac{\partial^{2}}{\partial r_{\ast}^{2}},\nonumber\\
& & \frac{\partial^{2}}{\partial
\theta^{2}}=\frac{\partial^{2}}{\partial
\theta_{\ast}^{2}}-\frac{rr_{H}''r_{H}(r-r_{H})-rr_{H}'^{2}(r-r_{H})+rr_{H}'^{2}r_{H}}
{2\kappa r_{H}^{2}(r-r_{H})^{2}}\frac{\partial}{\partial r_{\ast}}
\nonumber\\
& & -\frac{2rr_{H}'}{2\kappa
r_{H}(r-r_{H})}\frac{\partial^{2}}{\partial r_{\ast}\partial
\theta_{\ast}}+\frac{r^{2}r_{H}'^{2}}{4\kappa^{2}
r_{H}^{2}(r-r_{H})^{2}}\frac{\partial^{2}}{\partial r_{\ast}^{2}}.
\end{eqnarray*}
The Klein-Gordon Eq(\ref{c3}) can be rewritten as
\begin{eqnarray}
& &
\frac{r_{H}'^{2}+[1+2\kappa(r-r_{H})]\{-2rr_{H}(\dot{r}_{H}+ar_{H}'\sin\theta)+r_{H}^{2}[1+2\kappa(r-r_{H})]
(1-2ar\cos\theta-\frac{2m}{r})\}}{ 2\kappa
r_{H}^{2}(r-r_{H})[1+2\kappa(r-r_{H})]}
\nonumber\\
& & \times\frac{\partial^{2}\Phi}{\partial r_{\ast}^{2}}
+2\frac{\partial^{2}\Phi}{\partial v_{\ast}\partial
r_{\ast}}+\{\frac{1}{r_{H}^{2}r[1+2\kappa(r-r_{H})](r-r_{H})}\nonumber\\
& &
[2rr_{H}^{2}\dot{r}_{H}-2\dot{r}_{H}r_{H}r(r-r_{H})-r_{H}^{2}r(1-2ar\cos\theta-\frac{2m}{r})
+2ar_{H}^{2}r_{H}'r\sin\theta
\nonumber\\
& &
-2ar_{H}'r_{H}r(r-r_{H})\sin\theta-r_{H}r_{H}'(r-r_{H})\cot\theta-r_{H}''r_{H}(r-r_{H})+r_{H}'^{2}(r-r_{H})-r_{H}'^{2}r_{H}]\nonumber\\
& &
+\frac{2}{r}-\frac{2m}{r^{2}}-4a\cos\theta\}\frac{\partial\Phi}{\partial
r_{\ast}}
+\frac{2rr_{H}[1+2\kappa(r-r_{H})]a\sin\theta-2r_{H}'}{rr_{H}[1+2\kappa(r-r_{H})]}\frac{\partial^{2}\Phi}{\partial
r_{\ast}\partial\theta_{\ast}}\nonumber\\
& & +\frac{2\kappa(r-r_{H})}{1+2\kappa(r-r_{H})}[\frac{2}{r}
\frac{\partial\Phi}{\partial
v_{\ast}}+\frac{2a\sin\theta}{r}\frac{\partial\Phi}{\partial\theta_{\ast}}+\frac{1}{r^{2}}\frac{\partial^{2}\Phi}{\partial
\theta_{\ast}^{2}}\nonumber\\
& &    +\frac{\cot\theta}{r^{2}}\frac{\partial\Phi}{\partial
\theta_{\ast}}+\frac{1}{r^{2}\sin^{2}\theta}\frac{\partial^{2}\Phi}{\partial\varphi^{2}}-\mu^{2}\Phi]=0.\label{c4}
\end{eqnarray}
From the null hypersurface condition Eq(\ref{c1}) of Kinnersley
spacetime, the numerator of the coefficient on the term
$\frac{\partial^{2}\Phi}{\partial r_{\ast}^{2}}$ approaches to zero
at the horizon $r_{H}$. Therefore we can calculate the limit of the
coefficient using L'Hospital law. Assuming the limit is equal to an
undetermined constant $K$, so when $\kappa$ is selected as
\begin{equation}
\kappa=\frac{1}{2r_{H}}\frac{\frac{m}{r_{H}^{2}}-a\cos\theta-\frac{r_{H}'^{2}}{r_{H}^{3}}}
{\frac{m}{r_{H}^{2}}+a\cos\theta+\frac{r_{H}'^{2}}{2r_{H}^{3}}} +
\frac{1}{2r_{H}}\frac{\frac{m}{r_{H}^{2}}+a\cos\theta+\frac{r_{H}'^{2}}{2r_{H}^{3}}-\frac{1}{2r_{H}}}
{\frac{m}{r_{H}^{2}}+a\cos\theta+\frac{r_{H}'^{2}}{2r_{H}^{3}}},\label{8}
\end{equation}
we have $K=1$.

When $r$ approaches to $r_{H}$, the Klein-Gordon Eq(\ref{c4}) can be
transformed into
\begin{equation}
\frac{\partial^{2}\Phi}{\partial
r_{\ast}^{2}}+2\frac{\partial^{2}\Phi}{\partial v_{\ast}\partial
r_{\ast}}+B\frac{\partial^{2}\Phi}{\partial
r_{\ast}\partial\theta_{\ast}}-G\frac{\partial\Phi}{\partial
r_{\ast}}=0,\label{c5}
\end{equation}
where
\begin{eqnarray*}
B&=&2(a\sin\theta_{0}-\frac{r_{H}'}{r_{H}^{2}}),\nonumber\\
G&=&-\frac{1}{r_{H}}+\frac{2m}{r_{H}^{2}}+\frac{r_{H}'\cot\theta_{0}}{r_{H}^{2}}+
\frac{r_{H}''}{r_{H}^{2}}-\frac{r_{H}'^{2}}{r_{H}^{3}}+\frac{r_{H}'^{2}}{r_{H}^{2}}.
\end{eqnarray*}
Separate variables as following
\begin{equation}
\Phi=R(r_{\ast})\Theta(\theta_{\ast})e^{il\varphi-i\omega v_{\ast}},
\end{equation}
where $\omega$ is the energy of the Klein-Gordon particle, $l$ is
the projection of angular momentum on $\varphi$-axis, we can get
\begin{eqnarray}
& & \Theta'=\lambda\Theta,\nonumber\\
& & R''+(\lambda B-G-2i\omega)R'=0,\label{c6}
\end{eqnarray}
where the constant $\lambda$ is introduced by the separation of
variables.

Assuming
$\lambda=\lambda_{1}+i\lambda_{2},\lambda_{1},\lambda_{2}\in R$, the
Eqs(\ref{c6}) can be written as
\begin{eqnarray}
& & \Theta'=(\lambda_{1}+i\lambda_{2})\Theta,\nonumber\\
& & R''+[( \lambda_{1}+i\lambda_{2})B-G-2i\omega]R'=0,
\end{eqnarray}
and its solution is
\begin{eqnarray}
& & \Theta=c_{1}e^{(\lambda_{1}+i\lambda_{2})\theta_{\ast}},\nonumber\\
& & R=c_{2}e^{-[(
\lambda_{1}+i\lambda_{2})B-G-2i\omega]r_{\ast}}+c_{3},
\end{eqnarray}
where $c_{1},c_{2}$ and $c_{3}$ are integral constants,
$\theta_{\ast}$ is polar angle. Its radial ingoing and outgoing
components are respectively
\begin{eqnarray}
\psi_{in}&=&e^{-i\omega v_{\ast}},\nonumber\\
\psi_{out}&=&e^{-i\omega
v_{\ast}}e^{2i(\omega-\omega_{0})r_{\ast}}e^{(G-\lambda_{1}
B)r_{\ast}},
\end{eqnarray}
where
\begin{equation}
\omega_{0}=\frac{1}{2}\lambda_{2}B=\lambda_{2}(a\sin\theta_{0}-\frac{r_{H}'}{r_{H}^{2}}).
\end{equation}
The outgoing wave is rewritten as
\begin{equation}
\psi_{out}=e^{-i\omega
v_{\ast}}e^{2i(\omega-\omega_{0})r}e^{\bar{A}r}(\frac{r-r_{H}}{r_{H}})^{\frac{i(\omega-\omega_{0})}
{\kappa}}(\frac{r-r_{H}}{r_{H}})^{\frac{\bar{A}}{2\kappa}},
\end{equation}
where $\bar{A}=G-\lambda_{1} B$. It is obvious that the outgoing
wave is not analytical at the horizon $r_{H}$. Extending the
outgoing wave from outside to inside of the horizon analytically
through the negative half complex plane, we get
\begin{equation}
\tilde{\psi}_{out}=e^{-i\omega
v_{\ast}}e^{2i(\omega-\omega_{0})r_{\ast}}e^{\bar{{A}}r_{\ast}}e^{\frac{-i\pi\bar{A}}{2\kappa}}
e^{\frac{\pi(\omega-\omega_{0})}{\kappa}}.
\end{equation}
The scattering probability of outgoing wave at the horizon is
\begin{equation}
\left\vert\frac{\psi_{out}}{\tilde{\psi}_{out}}\right\vert^{2}=e^{-\frac{2\pi(\omega-\omega_{0})}{\kappa}}.
\end{equation}
According to explanation of Sannan\cite{14}, the outgoing wave has
black body spectrum
\begin{eqnarray}
N_{\omega}&=&\frac{1}{e^{\frac{\omega-\omega_{0}}{k_{B}T}}\pm1},\\
T&=&\frac{\kappa}{2\pi k_{B }}.
\end{eqnarray}
\section{Touch between a Kinnersley black hole and its Rindler
horizon}

If $m=0$ in Eq(\ref{c1}), we will get Rindler horizon equation as
\begin{equation}
2\dot{r}-(1-2ar\cos\theta)+2ar'\sin\theta-\frac{r'^{2}}{r^{2}}=0.\label{1}
\end{equation}
So the Klein-Gordon Eq(\ref{c3}) can be rewritten as
\begin{eqnarray}
& & 2\frac{\partial^{2}\Phi}{\partial v\partial
r}+\frac{2}{r}\frac{\partial\Phi}{\partial v}+(1-2ar\cos\theta)
\frac{\partial^{2}\Phi}{\partial r^{2}}+2a\sin\theta\frac{\partial^{2}\Phi}{\partial \theta\partial r}\nonumber\\
& & +\frac{2r-6ar^{2}\cos\theta}{r^{2}}\frac{\partial\Phi}{\partial
r}+ \frac{2a\sin\theta}{r}\frac{\partial\Phi}{\partial
\theta}+2a\cos\theta\frac{\partial\Phi}{\partial r}
+\frac{1}{r^{2}}\frac{\partial^{2}\Phi}{\partial \theta^{2}}\nonumber\\
& & +\frac{\cot\theta}{r^{2}}\frac{\partial\Phi}{\partial
\theta}+\frac{1}{r^{2}\sin^{2}\theta}\frac{\partial^{2}\Phi}{\partial
\varphi^{2}}-\mu^{2}\Phi=0.\label{3}
\end{eqnarray}
To study the equation outside the Rindler horizon $(r<r_{H})$, the
tortoise coordinate should be written as
\begin{eqnarray}
& &
r_{\ast}=r+\frac{1}{2\kappa(v_{0},\theta_{0})}\ln[\frac{r_{H}(v,\theta)-r}{r_{H}(v,\theta)}],
\nonumber\\
& &v_{\ast}=v-v_{0}, \theta_{\ast}=\theta-\theta_{0}.
\end{eqnarray}
The Klein-Gordon Eq(\ref{3}) can be rewritten as
\begin{eqnarray}
& &
\frac{r_{H}'^{2}+[1+2\kappa(r-r_{H})]\{-2rr_{H}(\dot{r}_{H}+ar_{H}'\sin\theta)+r_{H}^{2}[1+2\kappa(r-r_{H})]
(1-2ar\cos\theta)\}}{ 2\kappa
r_{H}^{2}(r-r_{H})[1+2\kappa(r-r_{H})]}
\nonumber\\
& & \times\frac{\partial^{2}\Phi}{\partial r_{\ast}^{2}}
+2\frac{\partial^{2}\Phi}{\partial v_{\ast}\partial
r_{\ast}}+\{\frac{1}{r_{H}^{2}r[1+2\kappa(r-r_{H})](r-r_{H})}\nonumber\\
& &
[2rr_{H}^{2}\dot{r}_{H}-2\dot{r}_{H}r_{H}r(r-r_{H})-r_{H}^{2}r(1-2ar\cos\theta)
+2ar_{H}^{2}r_{H}'r\sin\theta
\nonumber\\
& &
-2ar_{H}'r_{H}r(r-r_{H})\sin\theta-r_{H}r_{H}'(r-r_{H})\cot\theta-r_{H}''r_{H}(r-r_{H})+r_{H}'^{2}(r-r_{H})-r_{H}'^{2}r_{H}]\nonumber\\
& & +\frac{2}{r}-4a\cos\theta\}\frac{\partial\Phi}{\partial
r_{\ast}}
+\frac{2rr_{H}[1+2\kappa(r-r_{H})]a\sin\theta-2r_{H}'}{rr_{H}[1+2\kappa(r-r_{H})]}\frac{\partial^{2}\Phi}{\partial
r_{\ast}\partial\theta_{\ast}}\nonumber\\
& & +\frac{2\kappa(r-r_{H})}{1+2\kappa(r-r_{H})}[\frac{2}{r}
\frac{\partial\Phi}{\partial
v_{\ast}}+\frac{2a\sin\theta}{r}\frac{\partial\Phi}{\partial\theta_{\ast}}+\frac{1}{r^{2}}\frac{\partial^{2}\Phi}{\partial
\theta_{\ast}^{2}}\nonumber\\
& &    +\frac{\cot\theta}{r^{2}}\frac{\partial\Phi}{\partial
\theta_{\ast}}+\frac{1}{r^{2}\sin^{2}\theta}\frac{\partial^{2}\Phi}{\partial\varphi^{2}}-\mu^{2}\Phi]=0.\label{4}
\end{eqnarray}
Due to the Rindler horizon Eq(\ref{1}), the numerator of the
coefficient on the first term of above equation approaches to zero
at the Rindler horizon $r_{H}$. Therefore we can calculate the limit
of the coefficient using L'Hospital law. Following above procedure,
we obtain
\begin{equation}
\kappa=\frac{1}{2r_{H}}\frac{-a\cos\theta-\frac{r_{H}'^{2}}{r_{H}^{3}}}
{a\cos\theta+\frac{r_{H}'^{2}}{2r_{H}^{3}}} +
\frac{1}{2r_{H}}\frac{a\cos\theta+\frac{r_{H}'^{2}}{2r_{H}^{3}}-\frac{1}{2r_{H}}}
{a\cos\theta+\frac{r_{H}'^{2}}{2r_{H}^{3}}}.
\end{equation}
Then the equation can be transformed into
\begin{equation}
\frac{\partial^{2}\Phi}{\partial
r_{\ast}^{2}}+2\frac{\partial^{2}\Phi}{\partial v_{\ast}\partial
r_{\ast}}+B\frac{\partial^{2}\Phi}{\partial
r_{\ast}\partial\theta_{\ast}}-G\frac{\partial\Phi}{\partial
r_{\ast}}=0,\label{5}
\end{equation}
where
\begin{eqnarray*}
B&=&2(a\sin\theta_{0}-\frac{r_{H}'}{r_{H}^{2}}),\nonumber\\
G&=&-\frac{1}{r_{H}}+\frac{r_{H}'\cot\theta_{0}}{r_{H}^{2}}+
\frac{r_{H}''}{r_{H}^{2}}-\frac{r_{H}'^{2}}{r_{H}^{3}}+\frac{r_{H}'^{2}}{r_{H}^{2}}.
\end{eqnarray*}
Separating variables as
\begin{equation}
\Phi=R(r_{\ast})\Theta(\theta_{\ast})e^{il\varphi-i\omega v_{\ast}},
\end{equation}
we have radial solution of the equation
\begin{eqnarray}
\Psi_{in}&=&e^{-i\omega v_{\ast}},\nonumber\\
\Psi_{out}&=&e^{-i\omega v_{\ast}}e^{2i\omega
r_{\ast}}e^{Gr_{\ast}}.
\end{eqnarray}
The outgoing wave solution $\Psi_{out}$ can be rewritten as
\begin{equation}
\psi_{out}=e^{-i\omega v_{\ast}}e^{2i\omega
r}e^{Gr}(\frac{r_{H}-r}{r_{H}})^{\frac{i\omega}
{\kappa}}(\frac{r_{H}-r}{r_{H}})^{\frac{G}{2\kappa}},
\end{equation}
which is not analytical at the horizon. Extending the outgoing wave
from outside $(r<r_{H})$ to inside $(r>r_{H})$ of the horizon
analytically by turning $+\pi$ angle through the positive half
complex plane, we get
\begin{equation}
\tilde{\Psi}_{out}=e^{-i\omega v_{\ast}}e^{2i\omega
r_{\ast}}e^{Gr_{\ast}}e^{\frac{i\pi G}{2\kappa}}
e^{\frac{-\pi\omega}{\kappa}}.
\end{equation}
The relative scattering probability of outgoing wave at the horizon
is
\begin{equation}
\left\vert\frac{\Psi_{out}}{\tilde{\Psi}_{out}}\right\vert^{2}=e^{\frac{2\pi\omega}{\kappa}},
\end{equation}
so Hawking radiation spectrum is
\begin{eqnarray}
& &   N_{\omega}=\frac{1}{e^{\frac{\omega}{k_{B}T}}\pm1},\\
& &   T=\frac{-\kappa}{2\pi k_{B }}\nonumber\\
& &      =\frac{1}{2\pi k_{B
}}(\frac{1}{2r_{H}}\frac{a\cos\theta+\frac{r_{H}'^{2}}{r_{H}^{3}}}
{a\cos\theta+\frac{r_{H}'^{2}}{2r_{H}^{3}}} +
\frac{1}{2r_{H}}\frac{-a\cos\theta-\frac{r_{H}'^{2}}{2r_{H}^{3}}+\frac{1}{2r_{H}}}
{a\cos\theta+\frac{r_{H}'^{2}}{2r_{H}^{3}}}).\label{2}
\end{eqnarray}

When $a$ is constant, we will get Rindler horizon equation of an
observer which has uniformly accelerated rectilinear motion
\begin{equation}
-(1-2ar\cos\theta)+2ar'\sin\theta-\frac{r'^{2}}{r^{2}}=0.
\end{equation}
Its solution is rotating parabolic surface
\begin{equation}
r=\frac{1}{a(1+\cos\theta)}.\label{6}
\end{equation}
From Eq(\ref{2}), the Hawking temperature of the Rindler horizon is
constant
\begin{equation}
 T=\frac{-\kappa}{2\pi k_{B }}=\frac{a}{2\pi k_{B }}.
\end{equation}

Now we consider the touch between a Kinnersley black hole and its
Rindler horizon. While $r_{H}$ reaches extreme value, i.e.
$r_{H}'=0$, Eq(\ref{c1}) can be rewritten as
\begin{equation}
(2a\cos\theta_{1})r_{H}^{2}-(1-2\dot{r}_{H})r_{H}+2m=0,
\end{equation}
where $\theta_{1}$ is angle when $r_{H}$ reaches extreme value. The
solution is
\begin{equation}
r_{H}=\frac{(1-2\dot{r}_{H})\pm
\sqrt{(1-2\dot{r}_{H})^{2}-16ma\cos\theta_{1}}}{4a\cos\theta_{1}}.\label{c2}
\end{equation}
When $\dot{r}_{H}=0$ and $ma\ll1$, the solution can be rewritten as
\begin{equation}
r_{H1}\approx\frac{1}{2a\cos\theta_{1}},r_{H2}=2m,\label{7}
\end{equation}
obviously $r_{H2}$ belongs to black hole horizon. Comparing
Eq(\ref{6}) with Eq(\ref{7}), we find $r_{H1}$ belongs to Rindler
horizon and $\theta_{1}=0$. The extreme value of $r_{H}$ is
\begin{eqnarray}
r_{H1}&=&\frac{(1-2\dot{r}_{H1})+
\sqrt{(1-2\dot{r}_{H1})^{2}-16ma}}{4a},\nonumber\\
r_{H2}&=&\frac{(1-2\dot{r}_{H2})-
\sqrt{(1-2\dot{r}_{H2})^{2}-16ma}}{4a}.
\end{eqnarray}
As black hole horizon touches Rindler horizon, i.e. $r_{H1}=r_{H2}$,
we will have
\begin{eqnarray}
& &
(1-2\dot{r}_{H})^{2}=16ma,r_{H1}=r_{H2}=\frac{1-2\dot{r}_{H}}{4a},\\
& &   a=\frac{m}{r_{H1}^{2}}=\frac{m}{r_{H2}^{2}}.
\end{eqnarray}
Eq(\ref{8}) can be rewritten as
\begin{eqnarray}
& & T_{2}=\frac{\kappa_{2}}{2\pi
k_{B}}\nonumber\\
& & =\frac{1}{2\pi
k_{B}}\frac{1}{2r_{H2}}(\frac{\frac{m}{r_{H2}^{2}}-a}
{\frac{m}{r_{H2}^{2}}+a} +
\frac{\frac{m}{r_{H2}^{2}}+a-\frac{1}{2r_{H2}}}
{\frac{m}{r_{H2}^{2}}+a}),
\end{eqnarray}
and Eq(\ref{2}) can be rewritten as
\begin{eqnarray}
& & T_{1}=\frac{-\kappa_{1}}{2\pi
k_{B}}\nonumber\\
& & =\frac{1}{2\pi
k_{B}}\frac{1}{2r_{H1}}(\frac{\frac{m}{r_{H1}^{2}}-a}
{\frac{m}{r_{H1}^{2}}+a} +
\frac{\frac{m}{r_{H1}^{2}}+a-\frac{1}{2r_{H1}}}
{\frac{m}{r_{H1}^{2}}+a}).
\end{eqnarray}
So the temperature of touch point is
\begin{equation}
T=\frac{1}{2\pi
k_{B}}\frac{1}{2r_{H}}\frac{2a-\frac{1}{2r_{H}}}{2a}=-\frac{\dot{r}_{H}}{2\pi
k_{B}(1-2\dot{r}_{H})r_{H}}\label{c7}.
\end{equation}

\section{Conclusion and discussion}
Following Zhao's method, Hawking temperature of the Kinnersley
spacetime is investigated. It is found that the temperature relies
on both time and angle. The phenomenon of touch between black hole
horizon and Rindler horizon is similar to collision between two
black holes. When $\dot{r}_{H}=0$, the temperature of the touch
point in Eq(\ref{c7}) is equal to zero. It will violate the third
law of thermodynamics, so maybe the touch is impossible.

\section*{Acknowledgement}
One of the authors, Jian Yang, would like to thank Dr. Shiwei Zhou
and Dr. Xianming Liu for their helpful discussions . This research
is supported by the National Natural Science Foundation of
China(Grant Nos. 10773002,10875012,10875018).

\end{document}